\documentclass[11pt]{article}
\pdfoutput=1
\usepackage[utf8]{inputenc}
\usepackage{dsfont}
\usepackage{diagbox}
\usepackage{amsfonts}
\usepackage{mathtools}
\allowdisplaybreaks[4]        
\usepackage{amssymb}
\usepackage{euscript}     
\usepackage{braket}
\usepackage{starfont}
\usepackage{color,soul}         
\usepackage{braket}
\usepackage{tensor}        
\usepackage{amsthm}
\usepackage{graphicx}
\usepackage{slashed}
\usepackage{leftidx}
\usepackage{subfigure}
\usepackage{bbm}
\usepackage{indentfirst}

\definecolor{outerspace}{rgb}{0.25, 0.29, 0.3}
\definecolor{scarlet}{rgb}{1.0, 0.13, 0.0}
\usepackage[header,title,page,titletoc]{appendix}  
\definecolor{princetonorange}{rgb}{1.0, 0.56, 0.0}
\definecolor{WildStrawberry}{rgb}{1.0, 0.26, 0.64}
\definecolor{rossocorsa}{rgb}{0.83, 0.0, 0.0}
\definecolor{navyblue}{rgb}{0.0, 0.0, 0.5}
\usepackage[numbers,sort&compress]{natbib}  
\usepackage{float}
\usepackage[paper=a4paper,margin=1in]{geometry}
\newtheorem*{theorem*}{Theorem}
\newtheorem*{conjecture*}{Conjecture}

\usepackage{titlesec}

\pdfoutput=1
\usepackage{amsmath}
\usepackage{color}
\usepackage{amsfonts}
\usepackage{amssymb}
\usepackage{graphicx}
\usepackage{geometry}
\usepackage{amssymb,epsfig,subfigure}
\usepackage{amssymb}
\usepackage{hyperref}
\usepackage{comment}
\usepackage[font=footnotesize]{caption}
\usepackage[T1]{fontenc}
\usepackage[utf8]{inputenc}
\usepackage{latexsym}

\usepackage{cancel}

\usepackage[numbers,sort&compress]{natbib}  
\usepackage{float}

\usepackage{dsfont}
\usepackage{diagbox}

\usepackage{mathtools}
\allowdisplaybreaks[4]

\usepackage{euscript}     
\usepackage{braket}
\usepackage{starfont}
\usepackage{color,soul}         
\usepackage{tensor}        
\usepackage{amsthm}

\usepackage{slashed}
\usepackage{leftidx}
\usepackage{bbm}

\makeatletter
\renewcommand\section{\@startsection {section}{1}{\z@}%
                                 {-3.5ex \@plus -1ex \@minus -.2ex}
                                   {2.3ex \@plus.2ex}%
                                   {\normalfont\large\bfseries}}
\renewcommand\subsection{\@startsection{subsection}{2}{\z@}%
                                   {-3.25ex\@plus -1ex \@minus -.2ex}%
                                     {1.5ex \@plus .2ex}%
                                     {\normalfont\bfseries}}
\renewcommand\subsubsection{\@startsection{subsubsection}{3}{\z@}%
                                   {-3.25ex\@plus -1ex \@minus -.2ex}%
                                     {1.5ex \@plus .2ex}%
                                     {\normalfont\itshape}}
\makeatother

\def\pplogo{\vbox{\kern-\headheight\kern -29pt
\halign{##&##\hfil\cr&{\ppnumber}\cr\rule{0pt}{2.5ex}&\ppdate\cr}}}
\makeatletter
\def\ps@firstpage{\ps@empty \def\@oddhead{\hss\pplogo}%
  \let\@evenhead\@oddhead 
}
\def\maketitle{\par
 \begingroup
 \def\thefootnote{\fnsymbol{footnote}}
 \def\@makefnmark{\hbox{$^{\@thefnmark}$\hss}}
 \if@twocolumn
 \twocolumn[\@maketitle]
 \else \newpage
 \global\@topnum\z@ \@maketitle \fi\thispagestyle{firstpage}\@thanks
 \endgroup
 \setcounter{footnote}{0}
 \let\maketitle\relax
 \let\@maketitle\relax
 \gdef\@thanks{}\gdef\@author{}\gdef\@title{}\let\thanks\relax}
\makeatother

\numberwithin{equation}{section}

\newcommand\eea{\end{eqnarray}}
\newcommand\bea{\begin{eqnarray}}

\def\beq{\begin{equation}}
\def\eeq{\end{equation}}

\newcommand{\be}{\begin{equation}}
\newcommand{\ee}{\end{equation}}
\newcommand{\ba}{\begin{align}}
\newcommand{\ea}{\end{align}}
\newcommand{\bg}{\begin{gather}}
\newcommand{\eg}{\end{gather}}
\newcommand{\bseq}{\begin{subequations}}
\newcommand{\eseq}{\end{subequations}}

\usepackage{hyperref}
\hypersetup{
    colorlinks,
    citecolor=rossocorsa,
    filecolor=navyblue,
    linkcolor=navyblue,
    urlcolor=navyblue
}

\begin{document} 

\begin{titlepage}

\begin{center}

\phantom{ }
\vspace{3cm}

{\bf Proof of the universal density of charged states in QFT}
\vskip 0.5cm
Javier M. Mag\'an
\vskip 0.05in
\small{\textit{David Rittenhouse Laboratory, University of Pennsylvania}}
\vskip -.4cm
\small{\textit{ 209 S.33rd Street, Philadelphia, PA 19104, USA}}
\vskip -.10cm
\small{\textit{Theoretische Natuurkunde, Vrije Universiteit Brussel (VUB) }}
\vskip -.4cm
\small{\textit{ and The International Solvay Institutes}}
\vskip -.4cm
\small{\textit{ Pleinlaan 2, 1050 Brussels, Belgium}}

\begin{abstract}

We prove a recent conjecture by Harlow and Ooguri concerning a universal formula for the charged density of states in QFT at high energies for global symmetries associated with finite groups. An equivalent statement, based on the entropic order parameter associated with charged operators in the thermofield double state, was proven in a previous article by Casini, Huerta, Pontello, and the present author. Here we describe how the statement about the entropic order parameter arises, and how it gets transformed into the universal density of states. The use of the certainty principle, relating the entropic order and disorder parameters, is crucial for the proof. We remark that although the immediate application of this result concerns charged states, the origin and physics of such density can be understood by looking at the vacuum sector only. We also describe how these arguments lie at the origin of the so-called entropy equipartition in these type of systems, and how they generalize to QFT's on non-compact manifolds.

\end{abstract}
\end{center}

\small{\vspace{5 cm}\noindent magan@sas.upenn.edu\\}

\end{titlepage}

\setcounter{tocdepth}{2}

{\parskip = .4\baselineskip \tableofcontents}
\newpage

\section{Introduction to the conjecture and its different faces}

The aim of this article is simple to state. By using the recently introduced entropic order parameters for global symmetries \cite{Casini:2019kex,Magan:2020ake,Casini:2020rgj,Review}, we seek to prove a recent conjecture described in Ref. \cite{Harlow:2021trr} by Harlow and Ooguri, concerning the density of charged states in QFT. We also relate this conjecture with the so-called \emph{equipartition of entropy}, found in \cite{PhysRevB.98.041106} in the abelian case and more recently in \cite{Murciano:2020vgh,milekhin2021charge} in the non-abelian case, both in $1+1$ CFT's. By unraveling the common origin of all these results, we will generalize them in several ways.

Let us start with the conjecture. It says
\begin{conjecture*}\label{conj}\cite{Harlow:2021trr} 
In any quantum field theory with a finite-group global symmetry G, on any compact spatial manifold at sufficiently high energy the density of states of a each representation $r$ has the following form
\be
\rho_r(E)=\frac{d_r^2}{|G|}\,\rho(E)\,,\label{result}
\ee  
where $\rho(E)$ is the density of states at energy $E$.
\end{conjecture*}
We will comment on the issue of compactness and generalize to non-compact scenarios at the end of the article. For the moment we stick with the original form of the conjecture. Let us remark that this formula appeared earlier in the context of 2d CFT's in Ref. \cite{2020for}.

Looking at the previous formula one might naively think that something weird is happening at high energies, where entropy does not seem to be \emph{equipartitioned} through the different degrees of freedom. The naive reason is that the dimension of a representation is $d_r$ so a first guess would say that if entropy is to be equipartitioned, the prefactor should be $\frac{d_r}{\sum_r d_r}$, instead of $\frac{d_r^2}{|G|}=\frac{d_r^2}{\sum_r d_r^2}$. We will see that indeed entropy is equipartitioned, but there is a fraction of $\frac{d_r^2}{|G|}$ degrees of freedom per representation of the group $G$.

It will prove convenient to consider different faces of this conjecture. We first define the twist $\tau_g$ to be the global unitary representation of the element $g\in G$. As shown in \cite{Harlow:2021trr}, the previous conjecture is equivalent to the assertion that at sufficiently high energies the twisted partition function approaches
\be\label{factor}
Z(\beta, g)\equiv\textrm{Tr}(e^{-\beta H}\,\tau_g)\xrightarrow[\beta\rightarrow 0]{} Z(\beta, e)\,\delta_{g,e}\;,
\ee
where $e$ is the identity, $H$ is the Hamiltonian and $\delta_{g,g'}$ is the group Kronecker delta function. Equivalently, this says that the expectation value of the twist at sufficiently high temperature is $\delta_{g,e}$. The reason is the following. We can write the twisted partition functions in terms of the charged density of states as follows
\be 
Z(\beta, g)=\sum_r \frac{1}{d_r}\int dE\, \rho_r(E)\, \chi_r (g)\, e^{-\beta E}\;.
\ee
If the second version of the conjecture is true we obtain
\be 
\sum_r \frac{1}{d_r}\int dE\, \rho_r(E)\, \chi_r (g)\, e^{-\beta E}=\delta_{g,e}\int dE\, \rho (E)\, \, e^{-\beta E}\;.
\ee
Multiplying both sides of this relation by the character and using orthogonality of characters one arrives at the first version of the conjecture.

The third equivalent formulation of this conjecture concerns the structure of the thermal density matrix. Since the Hamiltonian is invariant under the symmetry, the thermal density matrix is invariant as well, and therefore it decomposes into blocks corresponding to the different representations. We can see this by defining the projectors\footnote{See \cite{Casini:2019kex,Casini:2020rgj} for extensive use of these projectors in the context of generic QFT's in any dimension.}
\be
P_r=\frac{d_r}{|G|} \sum_g \chi_r^*(g) \,\tau_g \,.\label{centry}
\ee
They are naturally labeled by the set of inequivalent irreducible representations of the group $G$. Using the orthogonality relation satisfied by the characters one can verify  that indeed
\be
P_r P_{r'}=\delta_{r,r'}\, P_{r}\,, \hspace{1cm}\sum_ r P_r=\mathbf{1}\, . 
\ee
These relations imply that the thermal density matrix decomposes as
\be 
\rho_{\beta}\equiv \frac{e^{-\beta H}}{Z(\beta,e)}= \oplus_r\,p_r \,\rho_{\beta}^r\;,
\ee
where the classical probabilities $p_r$ associated with each representations are just
\be 
p_r=\textrm{Tr}\left( \rho_{\beta}\,P_r\right) =\frac{d_r}{|G|} \sum_g \chi_r^*(g) \,\textrm{Tr}\left( \rho_{\beta}\, \tau_g\right) \;.
\ee
Given this generic structure, the third equivalent version of the conjecture then says that at high energies the thermal density matrix approaches
\be 
\rho_{\beta}\xrightarrow[\beta\rightarrow 0]{}  \oplus_r\,\frac{d_r^2}{G} \,\rho_{\beta}^r\;,
\ee
so that the thermal entropy reads
\be \label{entc}
S(\rho_{\beta})\equiv -\textrm{Tr}\left( \rho_{\beta}\log \rho_{\beta}\right) =\sum_r  \frac{d_r^2}{G}\, S(\rho_{\beta}^r)+S(p)\;,
\ee
where $S(p)$ is the classical Shannon entropy of $p_r=\frac{d_r^2}{G}$.

Finally, combining this with the first version, we can compute the universal behavior of Renyi entropies of $\rho_{\beta}^r$ at high temperatures, namely
\be \label{equi}
S_n (\rho_{\beta}^r)\equiv \frac{1}{1-n}\log \textrm{Tr} \left( \rho_{\beta}^r\right)^n =\frac{1}{1-n}\log \left[ \frac{\int dE \rho^r (E)\, e^{-\beta n r}}{\left( \int dE \rho^r (E)\, e^{-\beta  r} \right)^n }\right] \xrightarrow[\beta\rightarrow 0]{}\log \frac{d_r^2}{G}+S_n (\rho_{\beta})\;.
\ee
As it should, for $n=1$ this expression is consistent with the general form~(\ref{entc}). We also see that the corrections do not depend on $n$, suggesting certain maximally mixed density matrix behavior behind, which we confirm below. For abelian groups, for which $d_r=1$, the correction does not depend on $r$ either, and the analog behavior for a $U(1)$ group was dubbed entropy equipartition in \cite{PhysRevB.98.041106}. For non-abelian groups the correction depends on $r$, as also found in \cite{Murciano:2020vgh,milekhin2021charge} for WZW models. Holographic computations can be found in \cite{Rene1,Rene2}. As we commented above, and as we will see later in detail, this $r$-dependence, ultimately dependent on the $r$-dependence of the original version of the conjecture, should not be interpreted as invalidating the idea of entropy equipartition and the hidden maximally mixed density matrix.

To turn this conjecture into a theorem, we will explain how yet another version of the conjecture was proven recently in \cite{Casini:2019kex}, namely the saturation of the entropic order parameter at high temperatures in the Thermofield Double State (TFD). We will first explain why such saturation is another version of the theorem. This uses the certainty relation, to be reviewed below. We will then describe how such saturation is proved. This uses on one hand known tools extracted from the DHR approach to global symmetries in QFT \cite{Doplicher:1969tk,Doplicher:1969kp,Doplicher:1971wk,Doplicher:1973at}, and further developments in that direction by Longo and Rehren \cite{longo1989,Longo:1994xe}. On the other hand, we resort to the entropic order and disorder parameters, together with the certainty relation satisfied by them \cite{Casini:2019kex,Magan:2020ake,Casini:2020rgj}. 

The article is organized as follows. We will start by introducing the certainty relation in the next section. In the third section, we introduce aspects of the DHR approach to global symmetries. In section four we prove the conjecture. In section five we further comment on entropy equipartition and describe that the origin of the conjecture can already be found in the vacuum sector. We will end up with a discussion describing a heuristic but transparent path to understand the generality of this result, and generlize the result to QFT's on non-compact spatial manifolds.

\section{Certainties in Quantum Mechanics}

There is a famous and important relation between quantum entropies associated with commutant algebras. If $\mathcal{A}$ is an algebra and $\mathcal{A}'$ is its commutant, the relation simply says that
\be
S_{\omega}(\mathcal{A})=S_{\omega}(\mathcal{A}')\;,
\ee
in a globally pure state $\omega$. In this equation $S_{\omega}(\mathcal{A})$ is the quantum entropy of the algebra $\mathcal{A}$ in state $\omega$, a quantity that is unambiguously defined, see \cite{petz2007quantum}.

In a series of recent papers \cite{Casini:2019kex,Magan:2020ake,Casini:2020rgj}, a conceptually similar formula has been discovered for certain canonical non-commuting algebras $\mathcal{A}$ and $\tilde{\mathcal{A}}$. These pairs of algebras were called Complementary Observable Algebras (COA), due to their intimate relation to quantum complementarity. We now follow the generic presentation of Ref. \cite{Magan:2020ake}.

Instead of quantum entropies, the relation will be between relative entropies. For a finite-dimensional algebra $\mathcal{M}$, relative entropy is defined by
\be
S_{\mathcal{M}}\left(\omega\mid\phi\right):=\mathrm{Tr}_{\mathcal{M}}\left(\rho^\omega\left(\log\rho^\omega-\log\rho^\phi\right)\right)\,,\label{re_def}
\ee
where $\mathrm{Tr}_{\mathcal{M}}$ is the canonical trace on $\mathcal{M}$, and $\rho^\omega$, $\rho^\phi$ are the density matrices representing the underlying states, see \cite{ohya2004quantum}.

Now consider an inclusion of algebras $\mathcal{N}\subset\mathcal{M}$. In this scenario, there is typically a zoo of projections $\varepsilon : \mathcal{M}\rightarrow \mathcal{N}$, from the big algebra to the small algebra, see \cite{longo1989,Magan:2020ake} for concrete constructions of such a zoo in different scenarios. These are called conditional expectations and satisfy the bimodule property
\be
\hspace{-1mm} \varepsilon\left(n_{1}\,m\,n_{2}\right)=n_{1}\varepsilon\left(m\right)n_{2}\,,\hspace{3mm} \forall m\in\mathcal{M},\,\forall n_{1},n_{2}\in\mathcal{N}.\label{ce_def_prop}
\ee
Interestingly, given a state $\omega_{\mathcal{N}}$ in $\mathcal{N}$, and a conditional expectation $\varepsilon$, there is a canonical lift of $\omega_{\mathcal{N}}$ to a state in $\mathcal{M}$, namely the composition $\omega_{\mathcal{N}}\circ \varepsilon$. So for any state $\omega$ in the big algebra we can naturally define the following \emph{entropic order parameter}
\be 
S_{\mathcal{M}}\left(\omega\mid\omega\circ\varepsilon\right)\;.
\ee
If $\mathcal{M}=\mathcal{N}\vee \mathcal{A}$, that is, the algebra generated by $\mathcal{N}$ and $\mathcal{A}$, then the previous quantity is a refined measure of uncertainty of $\mathcal{A}$ that takes into account potential correlations between $\mathcal{N}$ and $\mathcal{A}$.

Given this structure, a canonical candidate for the COA, namely for $\tilde{\mathcal{A}}$, arises by taking commutants of the previous inclusion. This can be nicely characterized by a complementarity diagram
\begin{eqnarray}
\mathcal{M} & \overset{\varepsilon}{\longrightarrow} & \mathcal{N}\nonumber \\
\updownarrow\prime\! &  & \:\updownarrow\prime\\ \label{ecr_diagr}
\mathcal{M}' & \overset{\varepsilon'}{\longleftarrow} & \mathcal{N}'\,.\nonumber 
\end{eqnarray}
In this diagram, going vertically takes the algebras to its commutants, while horizontally in the arrow direction means we restrict to the target subalgebra. If $\varepsilon$ kills the algebra $\mathcal{A}\subset\mathcal{M}$, then the \textit{dual conditional expectation} $\varepsilon '$ kills $\tilde{\mathcal{A}}\subset\mathcal{N}'$. Notice that $\tilde{\mathcal{A}}$ does not commute with $\mathcal{A}$ by construction. Also notice that associated with the dual inclusion we can define a dual \emph{entropic disorder parameter}
\be 
S_{\mathcal{N}'}\left(\omega|\omega\circ\varepsilon'\right)\;.
\ee
As an example of a complementarity diagram, take as $\mathcal{M}$ the abelian algebra $\mathcal{X}$ generated by the position operator. Take also a conditional expectation that kills the full $\mathcal{M}=\mathcal{A} = \mathcal{X}$. Following the commutants we obtain
\bea
\mathcal{X} & \overset{\varepsilon}{\longrightarrow} & \mathds{1}\nonumber \\
\updownarrow\prime \!\! &  & \,\updownarrow\prime\\
\mathcal{X} & \overset{\varepsilon'}{\longleftarrow} & \mathcal{X}\vee\mathcal{P}\,.\nonumber 
\eea
We conclude that the COA associated with  $\mathcal{X}$ is the algebra generated by the momentum operator $\mathcal{P}$, as expected.

Given these dual inclusions associated with non-commuting COA, the main theorem of Ref. \cite{Magan:2020ake} states that given a globally pure state in $\mathcal{M}\vee\mathcal{N}'$, and a choice of conditional expectation $\varepsilon$, the dual conditional expectation $\varepsilon'$ can always be chosen to obtain the \emph{entropic certainty principle}
\be
S_{\mathcal{M}}\left(\omega|\omega\circ\varepsilon\right)+S_{\mathcal{N}'}\left(\omega|\omega\circ\varepsilon'\right)=\log\lambda \,, \label{cerp}
\ee
where $\lambda$ is the so-called Jones (or algebraic) index of the dual conditional expectations. Intuitively, the index measures the relative size of $\mathcal{N}$ in $\mathcal{M}$. Equivalently, it measures the size of $\mathcal{A}$. For subgroups, it is the usual index for the subgroup. But this notion can be generalized to include all algebraic scenarios. The first notion of the index was proposed by Jones in the context of inclusions of type $II_{1}$ subfactors \cite{Jones1983}. It was later noticed independently by Kosaki and Longo \cite{KOSAKI1986123,longo1989}, that the index was most naturally associated with a conditional expectation, and both were able to extend the definition to type $III$ algebras. For conditional expectations associated with averages over finite groups the index is the number of group elements $G$, see \cite{longo1989,Longo:1994xe,2001L,teruya} for considerations in different scenarios. 

An interesting feature of the certainty relation, that will be crucial for the proof of the conjecture, is that given the positivity of relative entropy, whenever the entropic order/disorder parameter saturates to the maximal value given by $\log \lambda$, then we know that the entropic disorder/order parameter will vanish, respectively.

The certainty relation was first found in QFT's with global symmetries \cite{Casini:2019kex}. It was then proven in generic inclusions of type I algebras in \cite{Magan:2020ake}, and shortly after extended to type III algebras \cite{Hollands:2020owv}. The entropic order parameters were used to characterize global symmetries in \cite{Casini:2019kex}, an approach which we use below. They were later put in a broader context as sensible order parameters for generalized symmetries in QFT in \cite{Casini:2020rgj}.

\section{Global symmetries and Cuntz algebras in QFT}

In this section we review the requiered tools from the DHR approach to global symmetries and superselection sectors in QFT \cite{Doplicher:1969tk,Doplicher:1969kp,Doplicher:1971wk,Doplicher:1973at,Doplicher:1990pn}, and some further developments in that direction by Longo and Rehren \cite{longo1989,Longo:1994xe}.

The DHR approach starts from the analysis of the observable algebra, which is the algebra invariant under the symmetry transformations. The theory then develops from the analysis of the category of localized endomorphisms of such algebra. This is a somewhat abstract approach, since in conventional uses and appearances of QFT's with global symmetries, local charged operators and unitary representations of the symmetry group play a prominent role. Nevertheless, even if the approach starts with the observable algebra, at the end of the day it is able to rederive all the structure of local charged oeprators. Indeed, the reconstruction theorem proved in Ref. \cite{Doplicher:1990pn} reconstructs the whole structure of charged operators and unitary symmetry representations from the categorical structure of the localized endomorphisms. Not only the existence of charged operators was known and derived in the DHR approach from more intrinsic entities, but also, as a byproduct, a better understanding of charged operators was indeed achieved. This understanding is what we seek to review here since it plays an important role below.

Usually, in conventional textbooks, a charged operator $\psi^r_i$ of a global symmetry is a local operator transforming under certain representation of the symmetry group, namely
\be 
\tau_g\,\psi^r_i\,\tau_g^{-1}=\sum_{j=1}^{d_r}\,r_{ij}(g)\,\psi^r_j\;,
\ee
where $r(g)$ is the matrix representing $g$ in representation $r$. The DHR approach shows that charged operators can be chosen to satisfy further convenient properties. In particular they can be chosen to be isometries and to form closed Hilbert spaces. To be concrete they can be chosen so that
\be 
\left( \psi^r_i\right)^{\dagger} \psi^r_j=\delta_{ij}\,,\hspace{1cm} \sum_{i=1}^{d_r} \psi^r_i\left( \psi^r_i\right)^{\dagger}=\mathbf{1}\;,
\ee
and then linear combinations of these charged operators can be seen to form a Hilbert space with inner product defined by the previous orthogonality. These types of relations later appeared in other different contexts, such as in the physics of anyons, see appendix E in the well-known Ref \cite{2006K} for example.

In the DHR approach these charged operators generate the endomorphisms of the observable algebra. There is one irreducible sector (endomorphim) per representation of the symmetry group. They read
\be
\rho_r:\mathcal{O}\rightarrow \rho_r \left( \mathcal{O}\right)\subset \mathcal{O} \,,\hspace{1cm}  \rho_r (\mathcal{O}) \equiv \sum_{i=1}^{d_r} \psi^r_i\,\mathcal{O}\,\left( \psi^r_i\right)^{\dagger}\;.
\ee
From this perspective, the charged operators are intertwiners from the identity representation to the charged superselection sector, in the usual sense of
\be 
\psi^r_i\, \mathcal{O}=\rho_r (\mathcal{O})\,\psi^r_i \;.
\ee
Once we have the intertwiners for the irreducible representations we can construct intertwiners for any reducible representation in the following way. Suppose we want to construct intertwiners for the endomorphim
\be\label{red} 
\rho\simeq \oplus_r\,N_r\,\rho_r\;,
\ee
where $\rho_r$ appears $N_r$ times. We just notice that in QFT, we can choose any partition of the identity for a given number of isometries. In our context choose $\omega_r^\alpha$, where $r$ runs over representations and $\alpha=1,\cdots,N_r$, such that
\be 
\left( \omega_r^\alpha\right) ^{\dagger}\omega_{r'}^{\alpha'}=\delta_{rr'}\delta_{\alpha\alpha'}\,,\hspace{1cm} \sum_{r\alpha}\omega_{r}^{\alpha}\left( \omega_r^\alpha\right) ^{\dagger}=\mathbf{1}\;.
\ee
The reducible endomorphim is then defined as
\be 
\rho (\mathcal{O})=\sum_{r\alpha}\omega_{r}^{\alpha}\rho_r (\mathcal{O})\left( \omega_r^\alpha\right) ^{\dagger}=\sum_{r\alpha i}\omega_{r}^{\alpha}\psi^r_i\,\mathcal{O}\,\left( \psi^r_i\right)^{\dagger}\left( \omega_r^\alpha\right) ^{\dagger}\equiv\sum_{r\alpha i}V^r_{i\alpha}\,\mathcal{O}\,\left( V^r_{i\alpha}\right) ^{\dagger} \;.
\ee
The intertwiners $V^r_{i\alpha}$ from the identity to the reducible representation are then
\be 
V^r_{i\alpha}\equiv\omega_r^\alpha\psi^r_i\,,\hspace{1cm} V^r_{i\alpha}\, \mathcal{O}=\rho (\mathcal{O})\,V^r_{i\alpha}\;,
\ee
and one can verify that also
\be 
\left( V^r_{i\alpha}\right) ^{\dagger}V^{r'}_{i'\alpha'}=\delta_{rr'}\delta_{\alpha\alpha'}\delta_{ii'}\hspace{1cm} \sum_{r\alpha i}V^{r'}_{i'\alpha'}\left( V^{r'}_{i'\alpha'}\right)^{\dagger}=\mathbf{1}\;.
\ee
Below, the crucial dominating case will be that of the regular endomorphism associated with the regular representation of the group. This is a reducible representation of dimension $G=\sum_{r}d_r^2$ for which $N_r=d_r$.

In general, the algebra generated by a set of operators $\psi_{\alpha}$, with $\alpha=1,\cdots, d$, satisfying 
\be 
\left( \psi_{\alpha}\right)^{\dagger}\psi_{\alpha'}=\delta_{\alpha\alpha'}\,,\hspace{1cm}\sum_{\alpha}^{d_r} \psi_{\alpha}\left( \psi_{\alpha}\right)^{\dagger}=\mathbf{1}\;,
\ee
is called a Cuntz algebra. It is an infinite dimensional algebra quite difficult to grasp. For our pourposes, and following \cite{longo1989} (see also \cite{Casini:2019kex}), the interesting thing is that these algebras contain very simple subalgebras of dimension $d^2$. Indeed, the algebra
\be 
(a)=\sum_{ij} a_{ij} V_r^i (V_r^j)^\dagger\;,
\ee
closes with a matrix multiplication for the coefficients 
\be
(a) (b)=(a\cdot b)\,.
\ee
Hence it is a finite subalgebra of the Cuntz algebra of matrices of $d\times d$. As we have shown above, we have one such algebra for any representation of the global symmetry group, even reducible ones. For the regular representation, it is an algebra of dimension $G\times G$.

\section{Universal density of charged states in QFT}
\label{chargedensity}

Having introduced the certainty principle and the Cuntz algebra of the regular representation, we are ready to prove the conjecture. We follow the construction in section $3.9$ of \cite{Casini:2019kex}. The thermofield double state of a duplicated QFT is defined as
\begin{equation}
\vert\textrm{TFD}\rangle =Z(\beta,e)^{-1/2}\sum\limits_{i}e^{-\beta E_{i}/2}\vert E_{i}^{R},E_{i}^{L}\rangle\,,
\end{equation}
where $L,R$ correspond to the Left/Right QFT. We will denote the left/right complete QFT algebras as $\mathcal{F}_L$ and $\mathcal{F}_R$. With complete we mean they include the invariant algebra and the charged operators as well. We will concentrate in the second version of the conjecture, namely the one concerning twisted partition functions~(\ref{factor}). These twisted partition functions can be written as expectation values of twist operators acting on the right (or left) system in the thermofield double state
\be 
\frac{Z(\beta, g)}{Z(\beta,e)}=\langle\textrm{TFD}\vert \,\tau_g\, \vert\textrm{TFD}\rangle\;.
\ee
In this context we can find entropic order and disorder parameters that sense aspects of the global symmetry. Notice we can construct the following complementarity diagram
\bea
{\cal O}_{L}\vee {\cal O}_{R}\vee \{I\} & \overset{\varepsilon}{\longrightarrow} &{\cal O}_{L}\vee {\cal O}_{R}\nonumber \\
\updownarrow\prime \!\! &  & \,\updownarrow\prime\\
1& \overset{\varepsilon'}{\longleftarrow} & \tau\,.\nonumber 
\eea
Let us describe the different pieces of this diagram. In the upper left part, ${\cal O}_{L},{\cal O}_{R}$ are the neutral (invariant under the symmetry) QFT left and right algebras. $I$ is the algebra of non-local intertwiners with one charged operator on one side and the other on the other side of the TFD. They are invariant under the common action of the symmetry group in both CFT's. We have one of these intertwiners per representation of the symmetry group, and representatives can be defined as
\be \label{inter}
I_r =\sum\limits_{i=1}^{d_r} \psi_{L i}^r (\psi_{R i}^r)^{\dagger}\;.
\ee
Using the regular representation, Ref. \cite{Casini:2020rgj} shows that representatives of these intertwiners can be chosen so that their fusion rules are the ones of the representations or characters of the symmetry group. Also, for the reader that finds the algebra ${\cal O}_{L}\vee {\cal O}_{R}\vee \{I\}$ somewhat arbitrary, we remark that such algebra is just the fixed point of the average over the symmetry group acting in the same way in both CFT's. In other words, if we call $E$ to such an average (a conditional expectation), then we have
\be \label{gloE}
E(\mathcal{F}_L\vee \mathcal{F}_R)={\cal O}_{L}\vee {\cal O}_{R}\vee \{I\}\;.
\ee
Coming back to the complementarity diagram, the conditional expectation $\varepsilon$ kills the intertwiner algebra, and projects the upper left part to the upper right part. 

The commutant of ${\cal O}_{L}\vee {\cal O}_{R}$ is the center of ${\cal O}_{L}\vee {\cal O}_{R}$, namely the algebra of invariant twits $\tau_c$. The invariant twists are defined irrespectively from the left or right global twists $\tau_g$ by averaging over conjugacy classes \cite{Casini:2019kex}. We have one invariant twist per conjugacy class $[c]$
\be 
\tau_c=\sum_{g\in [c]}\tau_g\;. 
\ee
The dual conditional expectation $\varepsilon'$  just kills all this algebra. Associated with this complementarity diagram there is a certainty relation
\be\label{cer}
S_{{\cal O}_{L}\vee {\cal O}_{R}\vee \{I\}}(\omega|\omega\circ \varepsilon)+S_{\tau}(\omega|\omega\circ \varepsilon')=\log G\;,
\ee
where the state $\omega$ is now the TFD state and $G$ is the number of group elements of the finite symmetry group. It is also the index $\lambda$ for the conditional expectations $E$ and $E'$, see \cite{Casini:2019kex,longo1989,Longo:1994xe,2001L}. To understand this we first notice that the index associated to an average over a group is $G$, which can be applied to this case in two ways
\be \label{Ig}
{\cal O}_{L}\vee {\cal O}_{R}\vee \{I\}\subset\mathcal{F}_L\vee \mathcal{F}_R\,,\hspace{1cm} \lambda=G\;,
\ee
and 
\be 
{\cal O}_{L}\subset\mathcal{F}_L\,,\hspace{1cm} \lambda=G\;,
\ee
and similarly for the right algebras. The reason the index for group averages is $\lambda=\sum_{r}\,d_r^2$ is because the group algebra is a direc sum of matrices of size $d_r^2$. Using the multiplicative behavior of the index \cite{longo1989} for tensor products, we see that
\be 
{\cal O}_{L}\vee {\cal O}_{R}\subset\mathcal{F}_L\vee \mathcal{F}_R\,,\hspace{1cm} \lambda=G^2\;.
\ee
Given~(\ref{Ig}), and the multiplicative behavior of the index under concatenation of conditional expectations \cite{longo1989} one concludes that
\be 
{\cal O}_{L}\vee {\cal O}_{R}\subset {\cal O}_{L}\vee {\cal O}_{R}\vee \{I\}\,,\hspace{1cm} \lambda=G\;.
\ee

Given the certainty relation~(\ref{cer}), as mentioned above, in case the entropic order parameter $S_{{\cal O}_{L}\vee {\cal O}_{R}\vee \{I\}}(\omega|\omega\circ \varepsilon)$ saturates to its maximal value $\log G$ at high energies, the certainty relation~(\ref{cer}) will force the entropic disorder parameter $S_{\tau}(\omega|\omega\circ \varepsilon')$ to vanish. For a relative entropy to vanish, the two states compared must be the same \cite{petz2007quantum}. This means that in such cases we have
\be 
\omega(\tau_g)=\omega\circ \varepsilon'(\tau_g)\,\rightarrow\,\omega(\tau_g)=\delta_{g,e}\,\rightarrow\, Z(\beta, g)=Z(\beta, e)\,\delta_{g,e}\,.
\ee
This shows that yet another version of the conjecture goes by asserting that the entropic order parameter saturates to its maximum value. This saturation was proven in \cite{Casini:2019kex}. We now describe that proof in detail.

As described in \cite{haag2012local}, thermal states and TFD can be defined through the KMS condition, namely the periodicity of correlation functions under shifts $\tau\rightarrow \tau +\beta$ of the imaginary time axis. Also, given an operator on the left system $V_{L}$, there is a mirror operator acting on the same way on the right system
\begin{equation}
V_{L}\,\vert\textrm{TFD}\rangle =J\,V_{R}\,J \,\vert\textrm{TFD}\rangle\;,
\end{equation}
where $J$ is an antiunitary operator.

To compute the entropic order parameter we use two steps. First, we use a key property of relative entropy, see \cite{petz2007quantum}. It says the following. Given an inclusion of algebras $\mathcal{N}\subset \mathcal{M}$, a conditional expectation $E: \mathcal{M}\rightarrow \mathcal{N}$ and two states $\omega$ and $\phi$ invariant under the conditional expectation
\be 
\omega=\omega\circ E\,,\hspace{1cm}       \phi=\phi\circ E\;,
\ee
then the relative satisfies
\be 
S_{\mathcal{N}}(\omega|\phi)=S_{\mathcal{M}}(\omega\circ E|\phi\circ E)\;.
\ee
In our scenario this means
\be 
S_{{\cal O}_{L}\vee {\cal O}_{R}\vee \{I\}}(\omega|\omega\circ \varepsilon)=S_{\mathcal{F}_L\vee \mathcal{F}_R}(\omega\circ E|\omega\circ \varepsilon\circ E)\;,
\ee
where $E$ is the average over the global group, whose action was defined in~(\ref{gloE}). This means we can uplift the computation to the complete algebra that includes the charged operators and still get the same result. In the second step, to compute the RHS of the previous equation, we use the monotonicity of relative entropy. The game is to choose the subalgebra $\mathcal{V}$ of $\mathcal{F}_L\vee \mathcal{F}_R$ that, allowing a precise computation, provides the best lower bound
\be \label{lowerV}
S_{\mathcal{F}_L\vee \mathcal{F}_R}(\omega\circ E|\omega\circ \varepsilon\circ E)\geq S_{\mathcal{V}}(\omega\circ E|\omega\circ \varepsilon\circ E)\;.
\ee
As for the subalgebra $\mathcal{V}$, Ref. \cite{Casini:2019kex} first chooses the subalgebras of the Cuntz algebras defined previously. For any representation, reducible or irreducible, of dimension $d$ we showed above we have charge creating operators $V^{i}$, with $i=1,\cdots, d$ satisfying the usual Cuntz algebra. These operators allow us to construct the algebra $\mathcal{A}$ as
\be
a\in \mathcal{A}\,,\hspace{1cm} a=\sum_{ij} a_{ij} V^i (V^j)^\dagger\;,
\ee
generated by the projectors $P_{ij}=V^i (V^j)^\dagger$. A convenient subalgebra $\mathcal{V}$ appears when we take the tensor product of the previous one in the left and right QFT's
\be 
\mathcal{V}\equiv\mathcal{A}_L\vee\mathcal{A}_R\;.
\ee
To find the relative entropy in the algebra $\mathcal{V}$, we need to compute all its correlation functions in the TFD state, namely
\begin{equation}\label{PrPl}
\rho^{\omega_{\textrm{TFD}}}_{jl,ik}=\langle\textrm{TFD}\vert\, P^R_{ij}\,P^L_{kl}\,\vert\textrm{TFD}\rangle\,.
\end{equation}
To maximize such correlation functions we choose in particular $P^L_{ij}=JP^R_{ij} J$. If $H_L$ and $H_R$ are the left and right Hamiltonians respectively, we can use the known relation
\be
e^{-\beta (H_{R}-H_{L})/2}V_{R}\vert\textrm{TFD}\rangle =J V_{R}^{\dagger} J \vert\textrm{TFD}\rangle\;,
\ee
see the Tomita-Takesaki theory decribed in \cite{haag2012local}. We arrive at
\begin{equation}
\rho^{\omega_{\textrm{TFD}}}_{jl,ik} =Z^{-1}\,\textrm{Tr}(e^{-\beta H_{R}/2}\,P^R_{ij}\,e^{-\beta H_{R}/2}\,(P^R_{kl})^{\dagger})=Z^{-1}\,\textrm{Tr}(e^{-\beta H_{R}}\,P^R_{ij}(-\beta/2)\,P^R_{lk})\;,
\end{equation}
where $P^R_{ij}(-\beta/2)\equiv e^{\beta H_{R}/2}\,P^R_{ij}\,e^{-\beta H_{R}/2}$ is the operator evolved in imaginary time. This expression makes easy the analysis of high and low temperatures. At high temperatures $\beta\rightarrow 0$ and we have $P^R_{ij}(-\beta/2)\rightarrow P^R_{ij}$. This implies
  \be
\rho^{\omega_{\textrm{TFD}}}_{jl,ik}\simeq  Z^{-1}\,\textrm{Tr}(e^{-\beta H_{R}}\, P^R_{ij} \,P^R_{lk}) = \delta_{jl} \, Z^{-1}\,\textrm{Tr}(e^{-\beta H_{R}} \,P^1_{ik})\,.
\ee
Invariance of the Gibbs ensemble under the symmetry groups says
\be 
Z^{-1}\,\textrm{Tr}(e^{-\beta H_{R}} P^1_{ik})=Z^{-1}\,\textrm{Tr}(e^{-\beta H_{R}} E( P^1_{ik}))=\frac{1}{d}\delta_{ik}\,,
\ee
so that
\be
\rho^{\omega_{\textrm{TFD}}}_{jl,ik}=d^{-1} \, \delta_{ik}\delta_{jl}\,.\label{md2}
\ee
This state is invariant under conjugation with any unitary transformation operator of the form 
\be
D\otimes D^*\,.
\ee
This is a pure state 
\be
S(\omega)=0\,,\label{twisted112}
\ee
and $\omega_{\textrm{TFD}}$ is maximally entangled between the $L$ and $R$ sides in charge space at sufficiently high temperatures. The second state, namely the state composed with the conditional expectation, can be computed in similar lines, see section $3.2$ in Ref. \cite{Casini:2019kex}. For a generic representation of dimension $d$, in which the irrep $r$ appears $N_r$ times, as in~(\ref{red}), so that $d=\sum_r \,N_r,d_r$, and defining the relative probability of a given representation to be
\be 
q_r\equiv\frac{N_r\,d_r}{d}\,,\hspace{1cm} \sum_r q_r=1\;,
\ee
the end result of the computation of Ref. \cite{Casini:2019kex} is that the relative entropy is given by
\be 
S_{\mathcal{V}}(\omega\circ E|\omega\circ \varepsilon\circ E)=-\sum_r q_r\log q_r+\sum_r q_r\log d_r^2\;.
\ee
To obtain the best lower bound in~(\ref{lowerV}) we have to maximize the previous result over the choice of representation. This is easily done. Such quantity is maximized whenever we have a direct sum of any number of regular representations for which $q_r=\frac{d^2_r}{G}$ and the bound becomes
\be 
S_{\mathcal{V}_{\text{regular}}}(\omega\circ E|\omega\circ \varepsilon\circ E)=\log G\;.
\ee
This is not only the best bound over the choice of subalgebras $\mathcal{V}$, but it is indeed tight since the entropic order parameter is bounded by above by the same amount. Combining all the different pieces
\be 
\log G\geq S_{{\cal O}_{L}\vee {\cal O}_{R}\vee \{I\}}(\omega|\omega\circ \varepsilon)=S_{\mathcal{F}_L\vee \mathcal{F}_R}(\omega\circ E|\omega\circ \varepsilon\circ E)\geq S_{\mathcal{V}_{\text{regular}}}(\omega\circ E|\omega\circ \varepsilon\circ E)\geq\log G\;,
\ee
we conclude that
\be 
S_{{\cal O}_{L}\vee {\cal O}_{R}\vee \{I\}}(\omega|\omega\circ \varepsilon)=\log G\;,
\ee
ending the proof of the conjecture in all its versions, which we now state as a theorem
\begin{theorem*}\label{theo}
In any quantum field theory with a finite-group global symmetry G, on any compact spatial manifold at sufficiently high energy the following statements are equivalent and true
\begin{itemize}

\item The density of states of a certain representation $r$ has the following form
\be
\rho_r(E)=\frac{d_r^2}{|G|}\,\rho(E)\,,
\ee  
where $\rho(E)$ is the density of states at energy $E$.

\item The twisted partition functions are given by
\be
Z(\beta, g)= Z(\beta, e)\,\delta_{g,e}\;.
\ee

\item The thermal state has the following decomposition
\be 
\rho_{\beta}\xrightarrow[\beta\rightarrow 0]{}  \oplus_r\,\frac{d_r^2}{G} \,\rho_{\beta}^r\;.
\ee

\item The thermal entropy has the following decomposition
\be 
S(\rho_{\beta}) =\sum_r  \frac{d_r^2}{G}\, S(\rho_{\beta}^r)+S(p_r)\;,
\ee
where 
\be 
S(\rho_{\beta}^r)=S(\rho_{\beta})+\log \frac{d_r^2}{G}\,,\hspace{1cm} p_r=\frac{d_r^2}{G}\;.
\ee

\item The entropic order parameter in the TFD saturates to its maximum value
\be 
S_{{\cal O}_{L}\vee {\cal O}_{R}\vee \{I\}}(\omega|\omega\circ \varepsilon)=\log G\;.
\ee

\item The entropic disorder parameter in the TFD vanishes
\be 
S_{\tau}(\omega|\omega\circ \varepsilon')=0\;.
\ee

\end{itemize}

\end{theorem*}

To finish this section, notice the behaviour at low temperatures is kind of the opposite.  The density matrix $Z^{-1}\,e^{-\beta H_{R}}$ tends to the projector into the vacuum state and
\begin{equation}
e^{-\beta H_{R}/2}V^i e^{-\beta H_{R}/2}\rightarrow 0\;,
\end{equation}
so that
\begin{equation}
e^{-\beta H_{R}/2}V^i (V^{j})^{\dagger}e^{-\beta H_{R}/2}\rightarrow \frac{1}{|G|}\delta_{ij}e^{-\beta H_{R}}\,.
\end{equation}
The correlation function~(\ref{PrPl}) then factorizes which means
\be
\rho^{\omega_{\textrm{TFD}}}_{jl,ik}=\frac{1}{|G|^{2}} \, \delta_{ij}\delta_{kl}\,.
\ee
This is the identity matrix in the subalgebra $(a)$, and it is of course invariant under independent left and right averages over the symmetry group. We conclude that at sufficiently low temperatures the state on the intertwiners is unperturbed by the conditional expectation and the associated relative entropy vanishes. The disorder parameter then saturates to its maximum value, as expected since in the vacuum we have $\langle\tau_g\rangle =1$, or equivalently
\be 
Z(\beta, g)\xrightarrow[\beta\rightarrow \infty]{} Z(\beta, e)\;.
\ee
Therefore, as $T$ goes from zero to infinity, the entropic order goes from zero to $\log |G|$ while the entropic disorder goes from $\log G$ to zero. There is a priori no critical temperature for the transition, and indeed it can be smooth. The entropy increases whenever the temperature crosses a threshold in which particles of a given representation become thermally excited.

\section{Equipartition of entropy in QFT, EE, and the vacuum sector}

One face of the above theorem concerns the sector structure of the thermal entropy in QFT's with global symmetries, namely the fourth bullet point in the previous theorem. This is a generalization of the \emph{equipartition of entropy}, found for vacuum EE for CFT's with $U(1)$ symmetry in 1+1 dimension in \cite{PhysRevB.98.041106} and in \cite{Murciano:2020vgh,milekhin2021charge} for WZW models. The saturation of the entropic order parameter generalizes those results to QFT's in any dimension at finite temperature and for any finite group. Still, the results of \cite{PhysRevB.98.041106,Murciano:2020vgh,milekhin2021charge} were originally framed for vacuum entanglement entropy, while the previous theorem has been framed as associated with thermal entropy of charged sectors. In this section we review the connection between thermal and vacuum sectors, following again Ref. \cite{Casini:2019kex}.

In the DHR approach to global symmetries, together with the reconstruction theorem, see \cite{Doplicher:1969tk,Doplicher:1969kp,Doplicher:1971wk,Doplicher:1973at,Doplicher:1990pn,longo1989,Longo:1994xe}, all the properties of the charged sector of the theory can be derived (or reconstructed) from the vacuum sector alone. The intuitive reason is that, locally, any charged excitation can be considered neutral in practice. The reason is that we cannot verify whether it is a charged operator or just one leg of a neutral intertwiner operator, as in~(\ref{inter}), where the second charged operator is localized far away. Therefore, at least on heuristic grounds, if we are seeking to understand the statistics of local charged operators in a certain state, we might equivalently seek to understand the statistics of intertwiner operators in the same state. The DHR approach and the reconstruction theorems make this loose intuition into a precise construction and theorem.

The previous observation means that the origin of the physics associated with the previous theorem should appear already in the vacuum sector alone. This was put on firm ground in Ref. \cite{Casini:2019kex} in the context of EE in the vacuum of QFT's with global symmetries. Without going into details, we just mention that several parts of the previous theorem, namely the saturation of the entropic order parameter to the maximum topological value $\log G$, the vanishing of the entropic disorder parameter, and the universal value of twisted partition functions (interpreted in the vacuum sector as expectation values of localized twists operators), were seen to occur when computing entanglement entropy in the vacuum and when taking the regularizing distance to zero. This is one of the main results in Ref. \cite{Casini:2019kex}, see also \cite{Hollands:2020owv} for a more rigorous derivation within the context of type III algebras. They constitute yet different equivalent versions of the previous theorem, which is now seen as being controlled by properties of the vacuum alone.

For example, the universal probabilities $d_r^2/G$, associated with each irreducible local sector in the vacuum, can be found in eq $3.72$ in Ref. \cite{Casini:2019kex} and the paragraph that follows that equation. We conclude that EE is equipartitioned in the same way in the vacuum (when the regularizing distance goes to zero) as the thermal entropy is in the thermal ensemble (when the temperature goes to infinity).

Moreover, apart from the previous proof, concerning Cuntz algebras and the regular representation, in the context of vacuum EE, there is a different, more straightforward, path to understanding such universality. When considering the symmetry contribution to the vacuum EE, the appropriate twisted partition functions now become expectation values of local twist operators. These local twist operators are defined to act as the global twists inside the appropriate region but then are rapidly smeared to act as the identity outside the region. This smearing has a natural non-zero width $\epsilon$, which is a physical regularizing distance for the contribution to entanglement entropy, see \cite{Casini:2019kex} for many specific examples. In this scenario, the universality we have been describing above simply follows from the fact that the expectation value of any \emph{bounded} operator in QFT goes to zero as its smearing becomes sharp. In the case of the localized twists, this occurs when the width of the smearing region vanishes. This limit is then associated with a high-temperature limit in the TFD. In this limit we again have $\langle \tau_g\rangle=\delta_{g,e}$, and the probability distribution $d_r^2/G$ follows universally, as mentioned above.

Interestingly, cases with sharp twists and continuous Lie groups can also be considered, see \cite{Casini:2019kex}. In these Lie group scenarios, the equipartition of entropy cannot happen exactly because there is an infinite number of group of elements and therefore an infinite number of representations. But it is important to remark that we can verify that all non-trivial representations are populated, or equivalently, the expectation values of projectors associated with any irreducible representation of the global symmetry group are non-zero. This just follows from the fact that the localized twist tends to the continuous group delta function for continuous groups, see \cite{Casini:2019kex}. This is consistent with the general grounds described in \cite{Casini:2020rgj,Review}. In particular, it is just a particular instance of the key statement that in QFT generalized symmetries come in pairs with precisely the same size. 

\section{Discussion}

The objective of this article was to turn the conjecture~(\ref{conj}) into a theorem~(\ref{theo}), and also describe all its different facets that had appeared previously in the literature, such as entropy equipartition and entropic order parameters. In the way, we have seen how the results of Ref.\cite{Casini:2019kex} generalize many important aspects to finite and continuous groups and any dimension.

The proof we have provided is somewhat involved. It mixes aspects of recent developments of information theory in QFT with not so widely known topics in the context of QFT's with global symmetries, namely the existence of Cuntz algebras for any representation of the symmetry group. One expects that a simpler argument should exist, perhaps heuristic to some extent but more aligned with the real origin of the phenomenon. In this final discussion section, we want to present what we believe is such an argument.

This argument is composed of two steps. The first step is something familiar. For any system, when increasing its temperature, we expect to asymptote to a maximally mixed density matrix, in which entropy or heat is equidistributed between all degrees of freedom. In the present case, if all representations begin to be populated at a certain critical temperature, we expect that this universal regime will be reached for temperatures much greater than the critical temperature.

The problem with this first step of the argument was described in similar terms in \cite{Harlow:2021trr} and in \cite{Murciano:2020vgh}. In the first reference, it was noticed that somehow the density of states is not partitioned as one could have expected for a maximally mixed density matrix. If we take into account that the dimension of each irreducible representation is $d_r$, one might expect a prefactor of the form $d_r/\sum_r d_r$, instead of the correct prefactor $d_r^2/G$. In the second reference, it was noticed that this strange dependence between entropy and dimensions of irreducible representations was going beyond the entropy equipartition found previously for abelian systems.

The second step of the argument solves those concerns. The key point is to notice that entropy is not equipartitioned between irreducible representations, it is equipartitioned between degrees of freedom. Whether those two types of equipartition are the same needs to be clarified, and indeed we argue now they are not. When we think in entropy equipartition between degrees of freedom, we immediately notice we have to take into account how reducible representations are populated as well, since, in QFT, the infinite local algebra will produce them ubiquitously. The same will happen in a many-body quantum lattice system, in the thermodynamic limit. So the question is what is the \emph{dominant} representation (the most populated one) when we are allowed to consider large products of charged operators. The answer to this question was described in  Ref.\cite{Casini:2019kex},\footnote{A similar argument recently appeared in \cite{Harlow:2021trr}.} where it was noticed that the regular representation is the only stable limit of the fusion of many representations,\footnote{In the TFD, as we proved above, we can then select charged operators in the regular representation together with the anticharges at the other side of the thermofield double to form neutral operators which are in a maximally entangled state at high temperature. Tracing out one of the QFT's leaves that regular algebra in a maximally mixed state.} a sort of fixpoint or central limit in the fusion. Relatedly, if we have a regular representation in a maximally mixed state, any tensor product with another state and another representation, with arbitrary probabilities for the irreducible representations, will give us again the fractional probabilities $\frac{d_r^2}{|G|}$ for the global state.

We conclude that both the universal density of states and the entropy decomposition found above, are just the natural outcome of the conventional entropy equipartition at high temperatures. The reason is that it is the regular representation the one dominating the ensemble of representations in the algebra of a many-body system with a finite global symmetry group.

From this new understanding, an interesting generalization of the previous theorem to QFT's on non-compact spatial manifolds follows. Let us describe this. For non-compact manifolds, the density of states is infinite, since it will naturally include a volume factor in its exponent. But we might expect the ratio between the total density of states and the charged density of states to show some universality as well. Indeed, say we start with a compact manifold with natural size $R$. The generalization is then
\be
\frac{\rho_r (E)}{\rho (E)}\xrightarrow[\frac{\beta}{R}\rightarrow 0]{}\frac{d_r^2}{G}\;.
\ee
In other words, if we take the non-compact limit, we do not need to heat to arbitrarily high temperatures. Any finite temperature will do the job within the infinite size limit. The reason is that, for any non-vanishing temperature, a Gibbs distribution has some non-vanishing probability of having the regular representation. If we have a big space this regular representation will appear again in many places. This is of course a heuristic argument. To make it precise we notice that in the non-compact scenario, the previous equation can be better completed as
\be
\frac{\rho_r (E)}{\rho (E)}\xrightarrow[\frac{\beta}{R}\rightarrow 0]{}\frac{d_r^2}{G}=\textrm{Tr}\left( \rho_{\beta}\,P_r\right) \;,
\ee
where $P_r$ are the projectors into the irreducible representations defined in the introduction. We can then consider the TFD state again. As in the previous theorem, if we prove the entropic disorder parameter vanishes then we are done. Equivalently, we can prove the entropic order parameter saturates to its maximum value $\log G$. In a TFD in which we do not make the temperature go to infinity, the previous argument does not hold, because the correlations will be smaller between the two sides. But if we have a large space, we can use the results of Ref. \cite{Casini:2020rgj}, section $3.3$. Basically, instead of one regular representation smeared over a certain region in both left and right QFT's, we can put many of those representations located far away from each other. This can be done when the spatial manifold grows in size. Then, for any non-vanishing correlation between the two sides of the TFD, and if we are allowed to locate a sufficient number of smeared operators well separate between each other, Ref. \cite{Casini:2020rgj} shows that the entropic order parameter saturates to its maximum value again. An illuminating example of this is the computation of entanglement entropy in ball-shaped regions performed in \cite{Casini:2019kex}. For CFT's this can be mapped to a thermal state at fixed temperature in hyperbolic space. Still, even at fixed (and indeed relatively low) temperature, since the volume of the hyperbolic space is infinite, one obtains by direct computation that the entropic order parameter satures to its maximum value.

\section*{Acknowledgements} 
The author is indebted to many discussions with H. Casini, M. Huerta and D. Pontello over the last years. We also wish to thank Pablo Bueno for comments on the draft. The work of J.M is supported by a DOE QuantISED grantDE-SC0020360 and the Simons Foundation It From Qubit collaboration (385592).

\newpage

\bibliography{EE}{}

\providecommand{\href}[2]{#2}\begingroup\raggedright\begin{thebibliography}{10}

\bibitem{Casini:2019kex}
H.~Casini, M.~Huerta, J.~M. Mag\'an, and D.~Pontello, ``{Entanglement entropy
  and superselection sectors. Part I. Global symmetries},''
  \href{http://dx.doi.org/10.1007/JHEP02(2020)014}{{\em JHEP} {\bfseries 02}
  (2020) 014}, \href{http://arxiv.org/abs/1905.10487}{{\ttfamily
  arXiv:1905.10487 [hep-th]}}.

\bibitem{Magan:2020ake}
J.~M. Magan and D.~Pontello, ``{Quantum Complementarity through Entropic
  Certainty Principles},''
  \href{http://dx.doi.org/10.1103/PhysRevA.103.012211}{{\em Phys. Rev. A}
  {\bfseries 103} no.~1, (2021) 012211},
  \href{http://arxiv.org/abs/2005.01760}{{\ttfamily arXiv:2005.01760
  [hep-th]}}.

\bibitem{Casini:2020rgj}
H.~Casini, M.~Huerta, J.~M. Magan, and D.~Pontello, ``{Entropic order
  parameters for the phases of QFT},''
  \href{http://dx.doi.org/10.1007/JHEP04(2021)277}{{\em JHEP} {\bfseries 04}
  (2021) 277}, \href{http://arxiv.org/abs/2008.11748}{{\ttfamily
  arXiv:2008.11748 [hep-th]}}.

\bibitem{Review}
H.~Casini and J.~M.~Magan, {\em On completeness and generalized symmetries in
  QFT}.
\newblock To appear soon in Mod. Phys. Lett. A.

\bibitem{Harlow:2021trr}
D.~Harlow and H.~Ooguri, ``{A universal formula for the density of states in
  theories with finite-group symmetry},''
  \href{http://arxiv.org/abs/2109.03838}{{\ttfamily arXiv:2109.03838
  [hep-th]}}.

\bibitem{PhysRevB.98.041106}
J.~C. Xavier, F.~C. Alcaraz, and G.~Sierra, ``Equipartition of the entanglement
  entropy,'' \href{http://dx.doi.org/10.1103/PhysRevB.98.041106}{{\em Phys.
  Rev. B} {\bfseries 98} (Jul, 2018) 041106}.
  \url{https://link.aps.org/doi/10.1103/PhysRevB.98.041106}.

\bibitem{Murciano:2020vgh}
S.~Murciano, G.~Di~Giulio, and P.~Calabrese, ``{Entanglement and symmetry
  resolution in two dimensional free quantum field theories},''
  \href{http://dx.doi.org/10.1007/JHEP08(2020)073}{{\em JHEP} {\bfseries 08}
  (2020) 073}, \href{http://arxiv.org/abs/2006.09069}{{\ttfamily
  arXiv:2006.09069 [hep-th]}}.

\bibitem{milekhin2021charge}
A.~Milekhin and A.~Tajdini, ``Charge fluctuation entropy of hawking radiation:
  a replica-free way to find large entropy,'' 2021.

\bibitem{2020for}
S.~Pal and Z.~Sun, ``High energy modular bootstrap, global symmetries and
  defects,'' \href{http://dx.doi.org/10.1007/jhep08(2020)064}{{\em Journal of
  High Energy Physics} {\bfseries 2020} no.~8, (Aug, 2020) }.
  \url{http://dx.doi.org/10.1007/JHEP08(2020)064}.

\bibitem{Rene1}
S.~Zhao, C.~Northe, and R.~Meyer, ``{Symmetry-resolved entanglement in
  AdS$_{3}$/CFT$_{2}$ coupled to U(1) Chern-Simons theory},''
  \href{http://dx.doi.org/10.1007/JHEP07(2021)030}{{\em JHEP} {\bfseries 07}
  (2021) 030}, \href{http://arxiv.org/abs/2012.11274}{{\ttfamily
  arXiv:2012.11274 [hep-th]}}.

\bibitem{Rene2}
K.~Weisenberger, S.~Zhao, C.~Northe, and R.~Meyer, ``{Symmetry-resolved
  entanglement for excited states and two entangling intervals in
  AdS${}_3$/CFT${}_2$},'' {\em accepted for publication in JHEP} (8, 2021) ,
  \href{http://arxiv.org/abs/2108.09210}{{\ttfamily arXiv:2108.09210
  [hep-th]}}.

\bibitem{Doplicher:1969tk}
S.~Doplicher, R.~Haag, and J.~E. Roberts, ``{Fields, observables and gauge
  transformations 1},''
\href{http://dx.doi.org/10.1007/BF01645267}{{\em Commun. Math. Phys.}
  {\bfseries 13} (1969) 1--23}.

\bibitem{Doplicher:1969kp}
S.~Doplicher, R.~Haag, and J.~E. Roberts, ``{Fields, observables and gauge
  transformations. 2.},''
\href{http://dx.doi.org/10.1007/BF01645674}{{\em Commun. Math. Phys.}
  {\bfseries 15} (1969) 173--200}.

\bibitem{Doplicher:1971wk}
S.~Doplicher, R.~Haag, and J.~E. Roberts, ``{Local observables and particle
  statistics. 1},'' \href{http://dx.doi.org/10.1007/BF01877742}{{\em Commun.
  Math. Phys.} {\bfseries 23} (1971) 199--230}.

\bibitem{Doplicher:1973at}
S.~Doplicher, R.~Haag, and J.~E. Roberts, ``{Local observables and particle
  statistics. 2},''
\href{http://dx.doi.org/10.1007/BF01646454}{{\em Commun. Math. Phys.}
  {\bfseries 35} (1974) 49--85}.

\bibitem{longo1989}
R.~Longo, ``Index of subfactors and statistics of quantum fields 1,'' {\em
  Comm. Math. Phys.} {\bfseries 126} no.~2, (1989) 217--247.
  \url{https://projecteuclid.org:443/euclid.cmp/1104179850}.

\bibitem{Longo:1994xe}
R.~Longo and K.-H. Rehren, ``{Nets of subfactors},''
  \href{http://dx.doi.org/10.1142/S0129055X95000232}{{\em Rev. Math. Phys.}
  {\bfseries 7} (1995) 567--598},
\href{http://arxiv.org/abs/hep-th/9411077}{{\ttfamily arXiv:hep-th/9411077
  [hep-th]}}.

\bibitem{petz2007quantum}
D.~Petz, {\em Quantum information theory and quantum statistics}.
\newblock Springer Science \& Business Media, 2007.

\bibitem{ohya2004quantum}
M.~Ohya and D.~Petz, {\em Quantum entropy and its use}.
\newblock Springer Science \& Business Media, 2004.

\bibitem{Jones1983}
V.~Jones, ``Index for subfactors.,'' {\em Inventiones mathematicae} {\bfseries
  72} (1983) 1--26. \url{http://eudml.org/doc/143011}.

\bibitem{KOSAKI1986123}
H.~Kosaki, ``Extension of jones' theory on index to arbitrary factors,''
  \href{http://dx.doi.org/https://doi.org/10.1016/0022-1236(86)90085-6}{{\em
  Journal of Functional Analysis} {\bfseries 66} no.~1, (1986) 123 -- 140}.
  \url{http://www.sciencedirect.com/science/article/pii/0022123686900856}.

\bibitem{2001L}
Y.~Kawahigashi, R.~Longo, and M.~Müger, ``Multi-interval subfactors and
  modularity¶of representations in conformal field theory,''
  \href{http://dx.doi.org/10.1007/pl00005565}{{\em Communications in
  Mathematical Physics} {\bfseries 219} no.~3, (Jun, 2001) 631–669}.
  \url{http://dx.doi.org/10.1007/PL00005565}.

\bibitem{teruya}
T.~Teruya, ``Index for von neumann algebras with finite-dimensional centers,''
  {\em Publ. Res. Inst. Math. Sci.} {\bfseries 28} (1992) 437–453.

\bibitem{Hollands:2020owv}
S.~Hollands, ``{Variational approach to relative entropies (with application to
  QFT)},'' \href{http://arxiv.org/abs/2009.05024}{{\ttfamily arXiv:2009.05024
  [quant-ph]}}.

\bibitem{Doplicher:1990pn}
S.~Doplicher and J.~E. Roberts, ``{Why there is a field algebra with a compact
  gauge group describing the superselection structure in particle physics},''
\href{http://dx.doi.org/10.1007/BF02097680}{{\em Commun. Math. Phys.}
  {\bfseries 131} (1990) 51--107}.

\bibitem{2006K}
A.~Kitaev, ``Anyons in an exactly solved model and beyond,''
  \href{http://dx.doi.org/10.1016/j.aop.2005.10.005}{{\em Annals of Physics}
  {\bfseries 321} no.~1, (Jan, 2006) 2–111}.
  \url{http://dx.doi.org/10.1016/j.aop.2005.10.005}.

\bibitem{haag2012local}
R.~Haag, {\em Local quantum physics: Fields, particles, algebras}.
\newblock Springer Science \& Business Media, 2012.

\end{thebibliography}\endgroup
\bibliographystyle{utphys}



\end{document}